\documentclass[trackchanges,modern]{aastex63}
\usepackage[T1]{fontenc}

\definecolor{DarkGreen}{rgb}{0.0,0.4,0.0}  
\newcommand{\B}[1]{{\color{blue}{#1}}}
\newcommand{\R}[1]{{\color{red}{#1}}}
\newcommand{\G}[1]{{\color{DarkGreen}{#1}}}
\newcommand{\M}[1]{{\color{magenta}{#1}}}
\newcommand{\C}[1]{{\color{cyan}{#1}}}
\usepackage[normalem]{ulem}
\usepackage{soul}
\usepackage{cancel}

\shorttitle{Saddle-like decay index}
\shortauthors{Luo \& Liu}

\received{January 23, 2022}
\revised{February 28, 2022}
\accepted{March 4, 2022}
\submitjournal{ApJ}

\begin{document}

\title{Where and how does a decay-index profile become saddle-like?}
\correspondingauthor{Rui Liu}
\email{rliu@ustc.edu.cn}

\author[0000-0002-2559-1295]{Runbin Luo}
\affiliation{CAS Key Laboratory of Geospace Environment, Department of Geophysics and Planetary Sciences, University of Science and Technology of China, Hefei 230026, China}

\author[0000-0003-4618-4979]{Rui Liu}
\affiliation{CAS Key Laboratory of Geospace Environment, Department of Geophysics and Planetary Sciences, University of Science and Technology of China, Hefei 230026, China}
\affiliation{CAS Center for Excellence in Comparative Planetology, University of Science and Technology of China, Hefei 230026, China}
\affiliation{Mengcheng National Geophysical Observatory, University of Science and Technology of China, Mengcheng 233500, China}

\begin{abstract}
The decay index of solar magnetic fields is known as an important parameter in regulating solar eruptions from the standpoint of the torus instability. In particular, a saddle-like profile of decay index, which hosts a local torus-stable regime at higher altitudes than where the decay index first exceeds the instability threshold, is found to be associated with some confined or two-step eruptions. To understand the occurrence of such a profile, we employed dipoles to emulate different kinds of photospheric flux distributions. Corroborated by observations of representative active regions (ARs), our major results are: 1) in bipolar configurations the critical height increases away from the AR center along the polarity inversion line (PIL) and its average is roughly half of the centroid distance between opposite polarities; 2) in quadrupolar configurations saddle-like profiles appear above the PIL when the two dipoles oriented in the same direction are significantly more separated in this direction than in the perpendicular direction, and when the two dipoles are oriented differently or have unequal fluxes; 3) saddle-like profiles in quadrupolar configurations are associated with magnetic skeletons such as a null point or a hyperbolic flux tube, and the role of such profiles in eruptions is anticipated to be double-edged if magnetic reconnection is involved. 
\end{abstract}


\section{Introduction} \label{sec:intro}

Solar flares and coronal mass ejections (CMEs) are the most energetic phenomena in solar corona, which are considered as the different manifestations of the same physical process \cite[]{ZhangJie2001,ZhangJie2004,Priest_EForbes2002,Shibata_Magara2011}. However, flares that are associated with no CMEs are known as ``confined flares'' or ``failed eruptions'' \cite[e.g.][]{Jihaisheng2003,Liu2014,Zhou2019,Mrozek2020}. Whether an eruption is successful or not is heavily dependent on the strapping force exerted by the overlying magnetic field on the eruptive structure \cite[e.g.,][]{TorokANDKliem2005,GuoY2010,Cheng2011,Zuccarello2014,Sun2015,MyersCE2015,Amari2018}, which is often a magnetic flux rope \cite[]{Vourlidas2013,LiuR2020}. A flux rope undergoes the torus instability \cite[]{Bateman1978,KliemANDTorok2006} when the magnitude of the overlying field decreases with height at a sufficiently steep rate, quantified by the decay index $n=-d\ln B/d\ln h$ \cite[]{Bateman1978} exceeding about 1.5; the corresponding height is referred to as the critical height. Nevertheless, one should bear in mind that the theoretical critical value of $n_\mathrm{cr}=1.5$ adopted in many studies as well as in this study is derived from a thin, toroidal current ring \citep{Bateman1978,KliemANDTorok2006}. In practice, $n_\mathrm{cr}$ typically scatters around in a range as large as [0.8, 2] for different theoretical considerations \cite[e.g.,][]{Demoulin&Aulanier2010,OlmedoANDZhang2010}, observational studies \cite[e.g.,][]{Jing2018,Duan2019}, numerical simulations \cite[e.g.,][]{TorokANDKliem2007,FanANDGibson2007}, and laboratory experiments \cite[e.g.,][]{MyersCE2015,Alt2021}. Obviously, $n_\mathrm{cr}$ is a function of the flux-rope parameters including the geometry and internal current profile, yet the discrepancy in different studies is also affected by where the decay index is computed \citep{Zuccarello2016,Alt2021}.
 
The CME-productivity of some active regions (ARs) are found to be effectively regulated by the decay index: ARs that produce mostly confined flares \citep[e.g., AR 12192;][]{LiuL2016} have significantly lower decay indices than those productive in eruptive flares \citep[e.g., AR 11158;][]{Sun2015}. Based on the pre-flare extrapolation of the coronal field, MFRs experiencing failed eruptions are found to stay within the torus-stable regime during the entire eruptive process \citep[e.g.,][]{WangHM2015,PengZ2019}. \cite{Wang2017} analyzed 60 two-ribbon flares and found that the critical heights for confined flares is higher than those for eruptive ones. 

A saddle-like decay-index profile is first reported by \cite{GuoY2010}, often exhibiting a local torus-stable regime sandwiched by two torus-unstable regimes. \cite{Wang2017} noticed such a profile for 16\% (14\%) of confined (eruptive) flares, with $n$ at the saddle bottom being significantly smaller in confined flares than that in eruptive ones. In another statistical study, \cite{Baumgartner2018} found saddle-like $n(h)$ profiles in 46\% of their sample of 44 M5.0-and-above flares. It is suggested that there might be an additional confining effect if the saddle bottom is below the critical value, which might be the case for some failed eruptions \cite[]{Wang2017,LiuL2018,Liting2019,Mitra2021} or two-step eruptions, in which the rising of the eruptive structure is temporarily slowed or even halted before developing into a CME \cite[e.g.,][]{Liu2007,Gosain2016,Chandra2017}. However, saddle-like profiles are also present in successful eruptions \cite[e.g., ][]{LiuChange2015,Wang2017,Baumgartner2018,Joshi2021}. In the data-driven numerical experiments by \cite{Inoue2018}, an MFR loses equilibrium in an torus-unstable regime but even further accelerate in the torus-stable regime of the saddle shape.

Saddle-like profiles tend to be present in complex, multipolar ARs \citep{Wang2017}. It is suggested that the associated magnetic configuration may possess two sufficiently different spatial scales \citep{Filippov2018,Kliem2021}. Here, we focus on the genesis of saddle-like decay-index profiles by idealized numeric experiments with magnetic dipoles. In the sections that follow, we present our methodology of using dipoles to emulate coronal background field in Section \ref{sec:M&D}; we then focus on where saddle-like profiles are present in the background field emulated by dipole and quadrupole field, respectively, and compare these results with the background field of a few representative active regions approximated by potential-field extrapolations in Section \ref{sec:Results}; we summarize our findings and discuss their implications in Section \ref{sec:C&D}.

\section{Methodology} \label{sec:M&D}
In view of the torus instability, an expanding toroidal flux ring can be constrained by an external poloidal field until its decay index $n\ge1.5$ \cite[]{KliemANDTorok2006}. In the corona, this external field is often approximated by a potential field extrapolated from the photospheric fields because, on one hand, it is difficult to distinguish the external field from that produced by the electric current flowing through a flux rope, and on the other hand, the potential field is usually perpendicular to the magnetic polarity inversion line (PIL, which separates magnetic fluxes of opposite polarities) along which a coronal flux rope is often aligned \citep[e.g.,][]{TorokANDKliem2007,FanANDGibson2007,LiuY2008,Demoulin&Aulanier2010}. The magnetic field produced by a magnetic dipole satisfies $\nabla \times \textbf{\textsl{B} = 0} $, which is strictly potential. We thus employ dipoles to emulate the coronal background field above ARs, and focus on the spatial distribution of decay index.

We selected five ARs with either a bipolar or quadrupolar configuration to compare with the dipole or quadrupole results, respectively. Magnetograms monitoring these ARs are obtained by the Helioseismic and Magnetic Imager \cite[HMI;][]{2014SoPh..289.3483H} on board the Solar Dynamics Observatory \cite[SDO;][]{2012SoPh..275....3P}, which was launched on 2010 February 11. The potential field to emulate the coronal background field above the ARs is extrapolated from the radial component of photospheric vector magnetograms by a Fourier transformation method \cite[]{1981A&A...100..197A}.

\section{Results} \label{sec:Results}
\subsection{Dipole fields} \label{subsec:DF}
A dipole field can be written as
\begin{equation} \label{eq:dip}
\mathbf{B}(\mathbf{r}) = \frac{3(\mathbf{m} \cdot \mathbf{r})\mathbf{r} }{r^5} - \frac{\mathbf{m}}{r^3},
\end{equation}
where $\mathbf{m}$ is the dipole moment and $\mathbf{r}$ the displacement vector. The plane at a height $h_0$ above the dipole moment, where the bipolar flux distribution (Fig.~\ref{fig:sim_quadrupole}a) shares similarities with a pair of sunspots with opposite polarities, is taken as a pseudo photosphere. In other words, the horizontal dipole $\mathbf{m}=m\,\mathbf{e}_x$ is placed at $(0,0,-h_0)$ in the Cartesian coordinate system. Right above the PIL, which is along the $y$-axis, the magnetic field can be written as
\begin{equation}
\mathbf{B}_\mathrm{PIL}(\mathbf{r}) = - \frac{m}{r^3}\,\mathbf{e}_x = -\frac{m}{[D^2+(h+h_0)^2]^\frac{3}{2}}\,\mathbf{e}_x, 
\end{equation} where $h$ is the height above the photosphere and $D$ is the linear distance from a point on the PIL to the origin. The decay index right above the PIL is then given as follows,
\begin{equation}\label{eq:dip_n}
n_\mathrm{PIL}=\frac{3 h (h+h_0)}{D^2+(h+h_0)^2}.
\end{equation}
The critical height $h_\mathrm{cr}$ is defined as the height where $n_\mathrm{PIL}=1.5$. In a dipole field,
\begin{equation}\label{eq:dip_hcrit_original}
h_\mathrm{cr}=(h_0^2 + D^2)^\frac{1}{2}.
\end{equation}

We define the centroid distance $d_c$ as the linear distance between the centroids of opposite polarities in the pseudo photosphere. The flux-weighted centroid of the positive polarity, e.g., is $(x_c,\,y_c)=(\int xB_z\,dA/\int B_z\,dA,\ \int yB_z\,dA/\int B_z\,dA)$, where $B_z>0$. We obtain $d_c$ in terms of the depth of the dipole $h_0$, i.e.,
\begin{equation}\label{eq:dip_d}
d_c=\pi h_0.
\end{equation}
$h_\mathrm{cr}$ can then be expressed in terms of $d_c$ to facilitate the comparison with observations,
\begin{equation}\label{eq:dip_hcrit}
h_\mathrm{cr}=\left(\frac{d_c^2}{\pi^2} + D^2\right)^\frac{1}{2}. 
\end{equation}
Taking the average critical height $\bar h_\mathrm{cr}$ over the PIL from $D= -0.7d_c$ to $0.7d_c$, we found
\begin{equation}\label{eq:dip_av_hcrit}
\bar h_\mathrm{cr} \simeq d_c/2.
\end{equation}

From Eq.~\ref{eq:dip_n} one can see that the decay index $n$ increases monotonically with the increasing height above the photosphere, which leaves no room for a saddle shaped $n(h)$ profile. Hence, below we investigate the decay index in quadrupole fields.

\subsection{Quadrupole fields} \label{subsec:QF}
In quadrupole fields, it becomes difficult to derive $h_\mathrm{cr}$ analytically. Hence, similar to \citet{2013ApJ...778..139S} and \citet{2016NatSR...634021L}, we placed two horizontal dipoles at the same depth below a Cartesian computation domain ($ -255.5 < x, y < 255.5$, $0 < z < 511 $). They yield together a potential field above the $z=0$ plane, which is again taken as the photosphere. Note in the 2D model of \cite{Filippov2018}, two dipoles are placed at different depths to generate the background field with two different spatial scales above the surface. Here for simplicity, $\mathbf{m_1}$'s flux distribution at the $z=0$ plane, termed ``BP1'' hereafter, is maintained constant, and $\mathbf{m}_1$ is always oriented in the $x$-direction. We adjust $\mathbf{m_2}$'s flux distribution at the $z=0$ plane, termed ``BP2'' hereafter, to produce different quadrupole configurations. The centroid distance $d_c$ of each individual dipole is the same by Eq.~\ref{eq:dip_d} and taken as the unit length in the following analysis. 

Two types of saddle-like profiles $n_\mathrm{sdl}(h)$ can be distinguished, based on whether the local maximum value at the saddle top $n_t$ is above $n_\mathrm{cr}$ and the local minimum value at the saddle bottom $n_b$ is below $n_\mathrm{cr}$. Saddle-like profiles satisfying $n_b < n_\mathrm{cr}< n_t$ (e.g., Fig \ref{fig:sim_quadrupole}g) only occur above the \emph{inter-bipole} PIL, i.e., the external PIL of each individual bipole but internal PIL of the quadrupole. This type of saddle-like profiles crosses the $n_\mathrm{cr}=1.5$ line for at least three times, which is denoted by $n_\mathrm{sdl}^\mathrm{m}(h)$ hereafter. Other saddle shapes satisfying either $n_b > n_\mathrm{cr}$ or $n_t < n_\mathrm{cr}$ (e.g., Fig.~\ref{fig:sim_quadrupole}(h \& i)) are found above the \emph{intra-bipole} PIL, mostly located close to the center of each individual bipole. They cross the $n_\mathrm{cr}=1.5$ line only once, which is denoted by $n_\mathrm{sdl}^\mathrm{s}(h)$ hereafter.

\subsubsection{Parameters regulating the occurrence of $n_\mathrm{sdl}(h)$} \label{subsubsec:where}
In observation, saddle-like $n(h)$ profiles are found exclusively in multipolar ARs \citep{Wang2017}. We simulate five types of quadrupole field configurations considering three parameters of BP2 relative to BP1, namely, position (Fig~\ref{fig:sim_quadrupole}(b,c,e)), direction (Fig~\ref{fig:sim_quadrupole}f), and intensity (Fig~\ref{fig:sim_quadrupole}d). The bipole direction, pointing from positive to negative polarity, is opposite to that of the corresponding dipole moment $\mathbf{m}$. The PIL segments above which the $n(h)$ profiles become saddle shaped are denoted by solid lines, while those above which $n(h)$ increases monotonically with height are denoted by dashed lines.

For simplicity we first consider two dipoles that are not only placed at the same depth but have the same orientation and strength. In comparison to a single dipole (Figure~\ref{fig:sim_quadrupole}a), the introduction of a second dipole aligned in the same direction produces saddle shaped $n(h)$ profiles above the major segments of the PIL (Figure~\ref{fig:sim_quadrupole}b). Since this configuration is similar to the breakout model \citep{Antiochos1999}, the resultant $n(h)$ breaks into two branches near the height where a coronal null point is present \citep[see the yellow curve in Fig.~\ref{fig:sim_quadrupole}g; see also Fig.3 in][]{TorokANDKliem2007}, which can be considered as a saddle shaped profile in the extreme, as $n$ is singular at the null point. In contrast, saddle shaped $n(h)$ are mostly absent above the PIL, i.e., $n$ increases monotonically with height, if BP1 and BP2 are aligned in the direction perpendicular to the bipole direction (Fig.~\ref{fig:sim_quadrupole}e).

Next, we relax the constraint on the orientation and strength of BP2. It seems that the broken of symmetry by either rotating (Fig.~\ref{fig:sim_quadrupole}f) or weaken BP2 relative to BP1 (Fig.~\ref{fig:sim_quadrupole}d) facilitates the production of saddle shaped $n(h)$. In particular, such profiles are preferentially found close to the weaker BP (Fig.~\ref{fig:sim_quadrupole}d). 

\subsubsection{Occurrence rate of $n_\mathrm{sdl}(h)$ above PIL} \label{subsubsec:QR}
To quantify the probability of encountering saddle-like $n(h)$ profiles above the PIL in quadrupole fields, we define the ``saddle occurrence rate'' as follows,
\begin{equation}\label{eq:saddle_rate}
R_\mathrm{sdl} = \frac{L_\mathrm{sdl}}{L},
\end{equation}
where $L$ is the length of the PIL from the center of BP1 to that of BP2, and $L_\mathrm{sdl}$ denotes the length of the PIL segments above which $n_\mathrm{sdl}(h)$ appears. Similarly, we may obtain the occurrence rate of $n_\mathrm{sdl}^\mathrm{m}(h)$ among saddle-shaped profiles, i.e.,
\begin{equation}\label{eq:msaddle_rate}
R_\mathrm{sdl}^\mathrm{m} = \frac{L_\mathrm{sdl}^\mathrm{m}}{L_\mathrm{sdl}}.
\end{equation} 
We set the height of the computational domain by the displacement of $\mathbf{m_2}$ relative to $\mathbf{m_1}$ in each individual quadrupole configurations. The result has been checked against those obtained in larger computational domains, but we found no qualitative difference.

First, two bipoles are set to be parallel to each other and have equal magnitude, but $\mathbf{m_2}$ is displaced with respect to $\mathbf{m_1}$ by $\Delta_x$ and $\Delta_y$ in the $X$ and $Y$ direction, respectively. The distributions of $R_\mathrm{sdl}$ and $R_\mathrm{sdl}^{\mathrm m}$ in this $\Delta_x$--$\Delta_y$ space are indicated by colors (Fig \ref{fig:sim_diff} (a) \& (b)). One can see that $\Delta_y/\Delta_x$ is an important parameter in regulating the occurrence rate of saddle shaped $n(h)$ above the PIL, i.e.,  $R_\mathrm{sdl}$ becomes significant ($>0.4$) when $\Delta_y/\Delta_x<1.1$ and $R_\mathrm{sdl}^\mathrm{m}$ starts to appear ($>0.1$) when $\Delta_y/\Delta_x<0.8$.

Next, as mentioned above, the angle $\theta$ between $\mathbf{m_1}$ and $\mathbf{m_2}$ can influence the generation of saddle shaped $n(h)$. We set up a quadrupole configuration with an initial $\Delta_x=0$ (e.g., Fig.~\ref{fig:sim_quadrupole}e) to ensure the absence of $n_\mathrm{sdl}(h)$, and then we change $\mathbf{m_2}$'s orientation to provide different $\theta$. In Fig.~\ref{fig:sim_diff}c the saddle occurrence rate $R_\mathrm{sdl}$ increases with $\theta$, when $\theta$ exceeds a threshold value ($\sim\,$90$^{\circ}$), but soon becomes saturated when $\theta\sim[130,145]^\circ$, and thence decrease gradually. At $\theta=180^\circ$, however, $R_\mathrm{sdl}$ suddenly drops to nearly 0, because this symmetric configuration possesses a single null point above the center. Only through the null point does the decay-index profile become extremely saddle-like (cf. the yellow curve in Fig.~\ref{fig:sim_quadrupole}g). Apparently, the threshold value depends on other parameters of the quadrupole configuration; e.g., the appearance of $n_\mathrm{sdl}^\mathrm{m}(h)$ requires a higher $\theta$ threshold than $n_\mathrm{sdl}^\mathrm{s}(h)$. 

Finally, we keep $\mathbf{m_1}$ constant and weaken $\mathbf{m_2}$ continuously to emulate asymmetric quadrupolar regions with unbalanced magnetic flux between two pairs of sunspots (Fig.~\ref{fig:sim_quadrupole}d). We divide PIL into two segments; namely, PIL1, which is close to BP1, and PIL2, which is close to BP2, based on the distance to the flux-weighted center of each pair of sunspots. The total saddle generation rate $R_\mathrm{sdl}$ above the entire PIL (black), $R_\mathrm{sdl,1}$ above PIL1 (blue), and $R_\mathrm{sdl,2}$ above PIL2 (red) are shown in Fig \ref{fig:sim_diff} (d), One can see that saddle-shaped $n(h)$ tends to appear above the PIL belonging to the weaker sunspot pair.

\subsection{Background fields of ARs} \label{subsec:observation}
Here we investigate a few selected active regions that have simple configurations, typically composed of either one or two pairs of major sunspots with opposite polarities, so that they can be emulated by either one or two dipoles. We further compare the decay index of the AR background fields that are approximated by a potential field with those emulated by dipoles.
\subsubsection{Bipolar active regions} \label{subsubsec:dipoleobs}
Three ARs, NOAA 12192, 11166, and 11428, which have two major sunspots of opposite polarities, are shown in Fig \ref{fig:obs_dipole}. The centroid distance $d_c$ is again given as the linear distance between the flux-weighted centroids of opposite polarities, and the midpoint is taken as the AR center (yellow triangles in Fig.~\ref{fig:obs_dipole}(a--c)). The color-coded decay index $n$ as a function of height above the PIL is plotted along the PIL (Fig.~\ref{fig:obs_dipole}(d--f)), whose origin is designated as the closest point to the AR center (yellow circles in Fig.~\ref{fig:obs_dipole}(a--c)). The decay index $n$ increases monotonically with increasing height above almost all segments of PIL. We took the average of $h_\mathrm{cr}$ over the PIL points whose arc length from the PIL origin along the PIL ranges from $-0.7d_c$ to $0.7d_c$ (thick curves in Fig.~\ref{fig:obs_dipole}(a--c)), and found that this average value, $\bar h_\mathrm{cr}$, is approximately $d_c/2$, consistent with Eq.~\ref{eq:dip_av_hcrit}.

To further compare with the dipole model, we show $h_\mathrm{cr}$ as a function of $D$, the linear distance from a PIL point to the AR center (Fig.~\ref{fig:obs_dipole}(g--i)). $h_\mathrm{cr}$ obtained for PIL segments to the north (south) of the PIL origin is denoted in blue (red) colors. That $h_\mathrm{cr}$ increases with the linear distance from the AR center suggests that magnetic structures located on the outskirts of a bipolar AR generally experience a stronger confinement imposed by the background field than those close to the AR center. An overall consistency between $h_\mathrm{cr}$ obtained from the potential-field approximation and those from the dipole field is demonstrated by Fig.~\ref{fig:obs_dipole}(g--i), in which the dashed curves are given by Eq.~\ref{eq:dip_hcrit_original}, with $h_0$ replaced by the critical height at the PIL origin obtained in the potential field.

\subsubsection{Quadrupolar active regions} \label{subsubsec:quadrupoleobs}
Here the coronal background fields of two exemplary ARs that consist of two major pairs of sunspots are compared with those idealized by a quadrupole field consisting of a pair of dipoles.

AR 11158 (Fig.~\ref{fig:obs_11158}a) can be emulated by a pair of dipoles which differ in both location and orientation as well as slightly in intensity (Fig.~\ref{fig:obs_11158}b). The $n(h)$ profiles plotted along the PIL (similar to Fig.~\ref{fig:obs_dipole}(d--f)) again demonstrate an overall similarity between the background field approximated by potential field and that by a pair of dipoles. We also plot $n(h)$ profiles at three specific locations in Fig \ref{fig:obs_11158}(g1--g3). Both $n(h)$ profiles at Location 1 and Location 3 can be categorized as $n_\mathrm{sdl}^\mathrm{s}(h)$, which crosses the $n=1.5$ line only once, but those at Location 2 can be categorized as $n_\mathrm{sdl}^\mathrm{m}(h)$, which crosses the $n=1.5$ line at least three times. A reversal in the transverse field component $\mathbf{B_t}$ and a strongly enhanced squashing factor $Q$ \citep{Titov2002} are found above Location 2 at similar heights in both the potential field and the quadrupole field (red and blue curves, respectively, in Fig.~\ref{fig:obs_11158}(h2 \& i2)), which will be discussed in \S \ref{subsec:role_of_sdl}.

Fig~\ref{fig:obs_11504_11505} shows a quadrupolar region composed of two bipolar ARs, NOAA 11504 and 11505. To emulate this quadrupolar field, we adopted two dipoles which have similar orientation but are relatively displaced only in the direction perpendicular to the dipole orientation. The $n(h)$ profiles show a similar trend, i.e., increasing monotonically with height, over the major PIL.

\section{Conclusions AND DISCUSSION} \label{sec:C&D}
The similarity in the distribution of $n(h)$ both along and above the PIL between the AR background fields approximated by potential-field extrapolations and those emulated by dipoles (\S\ref{subsec:observation}) suggests that the qualitative results that we learn from dipole and quadrupole fields (\S\ref{subsec:DF} \& \S\ref{subsec:QF}) can be applied to some simple ARs with caution. Hence below we summarize the major results obtained from this investigation using dipoles.

\subsection{Critical height in dipole fields}
In the dipole field, we found that the critical height averaged over above the major PIL in the core region $[-0.7d_c,\,0.7d_c]$ is approximately half of the centroid distance $d_c$ (Eq.~\ref{eq:dip_av_hcrit}). This is consistent with the statistics of two-ribbon flares by \citet[][their Eq.~(1)]{Wang2017}. Similarly, \citet{Kliem2014} found that $h_\mathrm{cr}$ is slightly below the half distance between two magnetic charges serving as the external field for a flux rope. These results generally imply that the bipolar ARs that have a large centroid distance between opposite polarities also have a large critical height, therefore providing strong confinement to underlying non-potential structures. This is reminiscent of the extremely large AR 12192, which produced 32 M- and 6 X-class flares but only one CME \citep{LiuL2016}. Eq.~\ref{eq:dip_hcrit} further implies that $h_\mathrm{cr}$ increases with increasing distance from the center of a bipolar region. However, this appears to contradict some observations, in which confined flares are located closer to the flux-weighted magnetic center than eruptive ones in the same AR \citep[e.g.,][]{WangANDZhang2007, Cheng2011,Baumgartner2018}. We note that the ARs in these studies are often far more complex than bipolar regions that we tried to emulate here with a dipole, also the overlying field of an eruptive structure located in the AR periphery must be modulated by the magnetic field of neighboring ARs (e.g., Figure~\ref{fig:obs_11504_11505}), which is seldom taken into account in previous studies.

\subsection{Occurrence condition of $n_\mathrm{sdl}(h)$ in quadrupole fields}
Placing two dipoles at the same depth below a plane taken as the photosphere, we found two types of saddle shaped $n(h)$ profiles in quadrupole fields: those cross the $n=1.5$ line multiple times and those cross the same line only once. The former, $n_\mathrm{sdl}^\mathrm{m}(h)$, is found only above the inter-bipole PIL between two bipoles, while the latter, $n_\mathrm{sdl}^\mathrm{s}(h)$, is close to and inside each bipole, i.e., mainly above the intra-bipole PIL.

We investigated three parameters regulating the generation of saddle shaped $n(h)$, namely, the relative location, orientation, and magnitude of the two dipoles.
\begin{itemize}
\item For two parallel dipoles, the rate of $n_\mathrm{sdl}(h)$ along the PIL becomes significant ($R_\mathrm{sdl}>0.4$) when their relative displacement along the dipole direction is larger than that perpendicular to the dipole direction; $n_\mathrm{sdl}^\mathrm{m}(h)$ appears when the former displacement is significantly larger than the latter. 
\item $n_\mathrm{sdl}(h)$ appears more frequently above the PIL with larger angle between the two dipoles.
\item $n_\mathrm{sdl}(h)$ tends to appear above the PIL segments closer to the bipole that is weaker in terms of unsigned flux.
\end{itemize} 

\subsection{Can $n_\mathrm{sdl}(h)$ provide additional constraint for eruptive structure?}\label{subsec:role_of_sdl}
The saddle shaped profile $n_\mathrm{sdl}(h)$, especially $n_\mathrm{sdl}^\mathrm{m}(h)$, provides a zone of stability for an eruptive structure when it rises to the height corresponding to the bottom of the saddle where $n<n_\mathrm{cr}$. It appears to be the case in some confined flares \cite[e.g.,][]{LiuL2018} or two-step eruptions \cite[e.g.,][]{Gosain2016}. However, it is also found that eruptions tend to occur at the ``external PIL'' of bipolar ARs \citep{Yardley2018}, which suggests that remote active regions might also play a role,  or at the ``collisional PIL'' between two pairs of sunspots within quadrupolar ARs \citep{Chintzoglou2019}. In both occasions, a saddle-like profile is likely present, from our experience of quadrupole fields (\S\ref{subsec:QF}). 
	
Further investigating the characteristics of magnetic field, we found that in quadrupole fields, the saddle-like decay-index profile is associated with a strong variation of magnetic connectivity; it is hence also associated with a significant rotation of the transverse field component $\mathbf{B_t}$, whose magnitude $B_t=\sqrt{B_x^2+B_y^2}$, with respect to the photospheric PIL \citep[see also][]{Filippov2020}. In the example shown in Fig~\ref{fig:Sdl&HFT}, two dipoles are relatively displaced but have the same orientation, strength and depth (Fig~\ref{fig:Sdl&HFT}a). Two quasi-separatrix layers \citep[QSLs;][]{Titov2002} can be seen as the isosurfaces of logarithmic squashing factor, $\log_{10}Q=3.5$. Their footprints at the pseudo photosphere are shown as the blue and red lines in the map of signed $\log_{10}Q$ (Fig~\ref{fig:Sdl&HFT}b). Such two QSLs intersecting each other are collectively known as a hyperbolic flux tube (HFT), whose intersections continuously transform from one end to the other \citep{Titov2002}, and in the center the intersection exhibits a typical X-shaped morphology (Fig~\ref{fig:Sdl&HFT}c). In the cross section above the inter-bipole PIL (marked by the thick curve in Fig~\ref{fig:Sdl&HFT}(a \& b)) where $n_\mathrm{sdl}^\mathrm{m}(h)$ is present (Fig~\ref{fig:Sdl&HFT}(d)), one can see that with increasing height the decay index changes from over $n_\mathrm{cr}=1.5$ (reddish) to below it (blueish) near where the two QSLs intersect to give an X shape. Near the X shape, magnetic field lines are roughly parallel to the PIL (magenta); but below and above, they become increasingly perpendicular to the PIL (yellowish and greenish, respectively). Right above the center of the PIL (Fig~\ref{fig:Sdl&HFT}(e--g)), one can see that the peak of $\log_{10} Q$ corresponds to where $n$ decreases most rapidly and to where the field lines rotate with respect to the PIL (shaded bar). 

For the background field of the representative AR 11158 as approximated by both potential field and two dipoles (Figure~\ref{fig:obs_11158}), we see similar height variation of various parameters, i.e., a reversal of the angle $\varphi$ between $\mathbf{B_t}$ and PIL is associated with strongly enhanced $Q$, along the PIL segment between sunspot pairs, where $n_\mathrm{sdl}^\mathrm{m}(h)$ is present (e.g., Location 2 in Figure~\ref{fig:obs_11158} and panels (g2, h2, j2)). Along the PIL segments within the sunspot pairs (e.g., Locations 3 in Figure~\ref{fig:obs_11158}), where only $n_\mathrm{sdl}^\mathrm{s}(h)$ is present, field lines rotate within a much smaller range without an orientation reversal with respect to PIL, despite the enhanced $Q$. An exception is found at Location 1, where $n(h)$ is not exactly saddle-like in the potential-field extrapolation, despite that the quadrupole field gives an untypical $n_\mathrm{sdl}^\mathrm{s}(h)$ profile. 
Location 1 is located at the intra-bipole PIL of the sunspot pair in the northwest of AR 11158. Compared with its southeastern counterpart, this pair is quite sprawled and its opposite polarities are highly sheared, which may contribute to the poor performance of the dipole in this region. Overall, unlike the relatively gradual rotation of the field lines, the drastic changes in magnetic connectivity as measured by the mapping of field-line footpoints, i.e., the squashing factor $Q$, can obviously serve as a more sensitive indicator for the presence of a saddle-shaped decay-index profile.

Imagine that initially a low-lying MFR aligned along the PIL is strapped by the downward Lorentz force between the axial current $\mathbf{J_a}$ of the MFR and the transverse component $\mathbf{B_t}$ of the overlying field, i.e., $F=J_a\,B_t\,\sin\varphi$, where $\varphi$ is the angle between $\mathbf{J_a}$ and $\mathbf{B_t}$ and $\sin\varphi<0$ in the low atmosphere. One may argue that the Lorentz force would turn upward, if the MFR somehow rose into the region where the $\mathbf{B_t}$ orientation is reversed with respect to the PIL (Figure~\ref{fig:Sdl&HFT}f). However, without magnetic reconnection to open up the overlying field, the rising MFR would just push the overlying field upward.

HFT is a preferential site for magnetic reconnection \cite[]{Titov2002,Titov2003}. When a rising MFR impacts the overlying field, the supposed reconnection between them would reduce the overlying flux and therefore weaken the constraining force, which facilitates the MFR's eruption, similar to the role of the null-point reconnection in the breakout model \cite[]{Antiochos1999}. On the other hand, the reconnection would shed magnetic twist from the MFR \cite[e.g.,][]{XueZK2016,Wyper2017,HuangZH2018,YangLH2019}, causing a redistribution of magnetic helicity over the larger system, and even the (partial) disintegration of the MFR \cite[e.g.,][]{LiuR2018,YanXL2020,ChenJL2021}, which hampers the MFR's eruption.

\begin{figure*}
\plotone{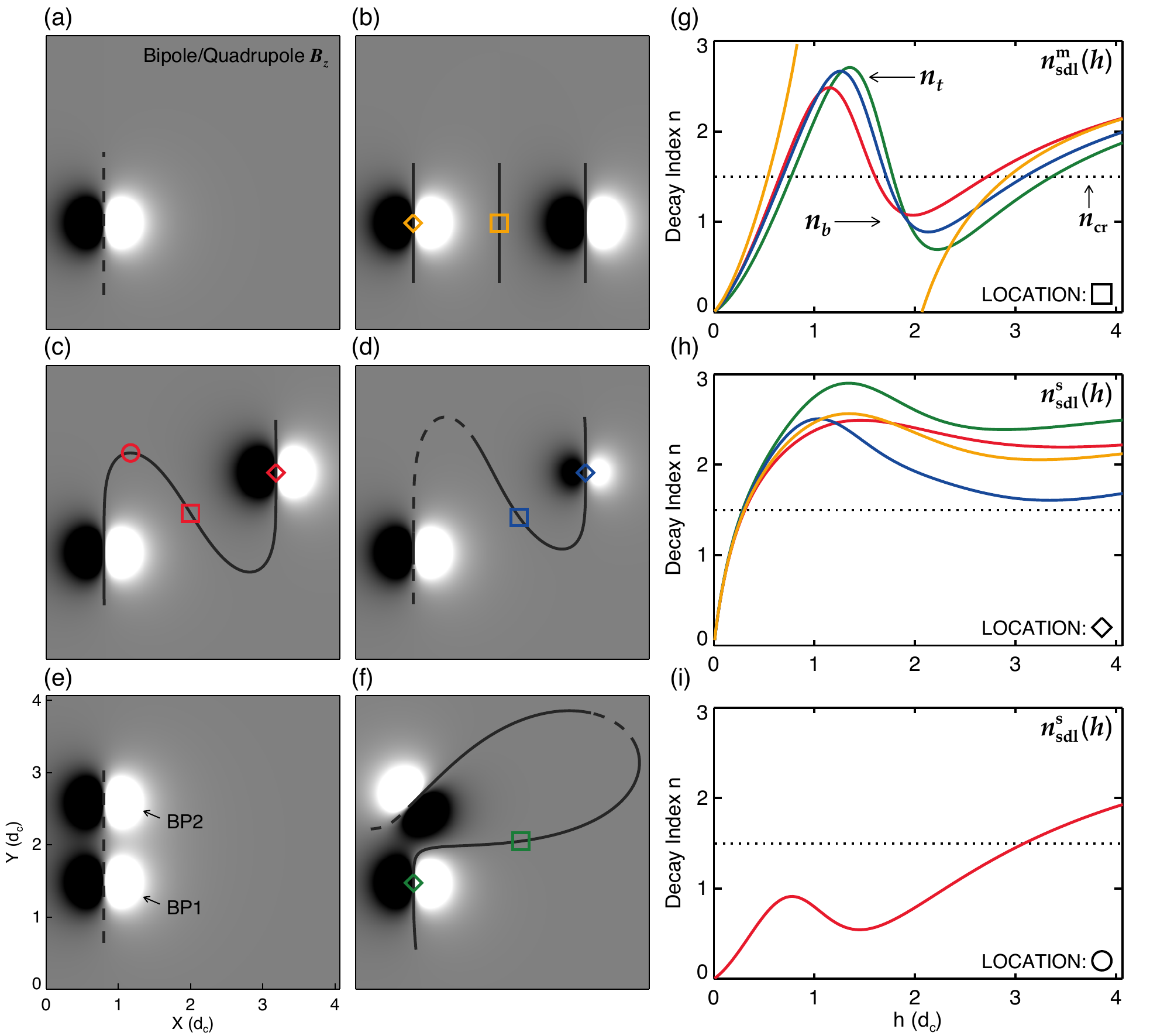}
\caption{Typical decay index profiles $n(h)$ above the PIL of different magnetic configurations. The photospheric $B_z$ map of (a) bipolar and (b--f) quadrupolar configurations consisting of two dipoles, BP1 and BP2. The two dipoles can be displaced only in the dipole direction (b) or only in the perpendicular direction (e)  or in both directions (c). (d) is similar to (c) except that BP2 is weaker in intensity. (f) is similar to (e) except that BP2 is rotated with respect to BP1. PIL segments associated with (without) saddle-like $n(h)$ are denoted by solid (dashed) curves. Panel (g) shows saddle-like decay-index profiles $n_\mathrm{sdl}^\mathrm{m}(h)$, which cross $n=1.5$ multiple times above the PIL points marked by squares. Panels (h) and (i) show saddle-like decay-index profiles  $n_\mathrm{sdl}^\mathrm{s}(h)$, which cross $n=1.5$ only once. The difference is that the saddle bottom is above (h) or below (i) $n=1.5$, corresponding to PIL points marked by diamonds or circles. The height $h$ is in units of the centroid distance $d_c$ of the dipole. \label{fig:sim_quadrupole}}
\end{figure*}

\begin{figure*}
\centering
\includegraphics[width=0.9\linewidth]{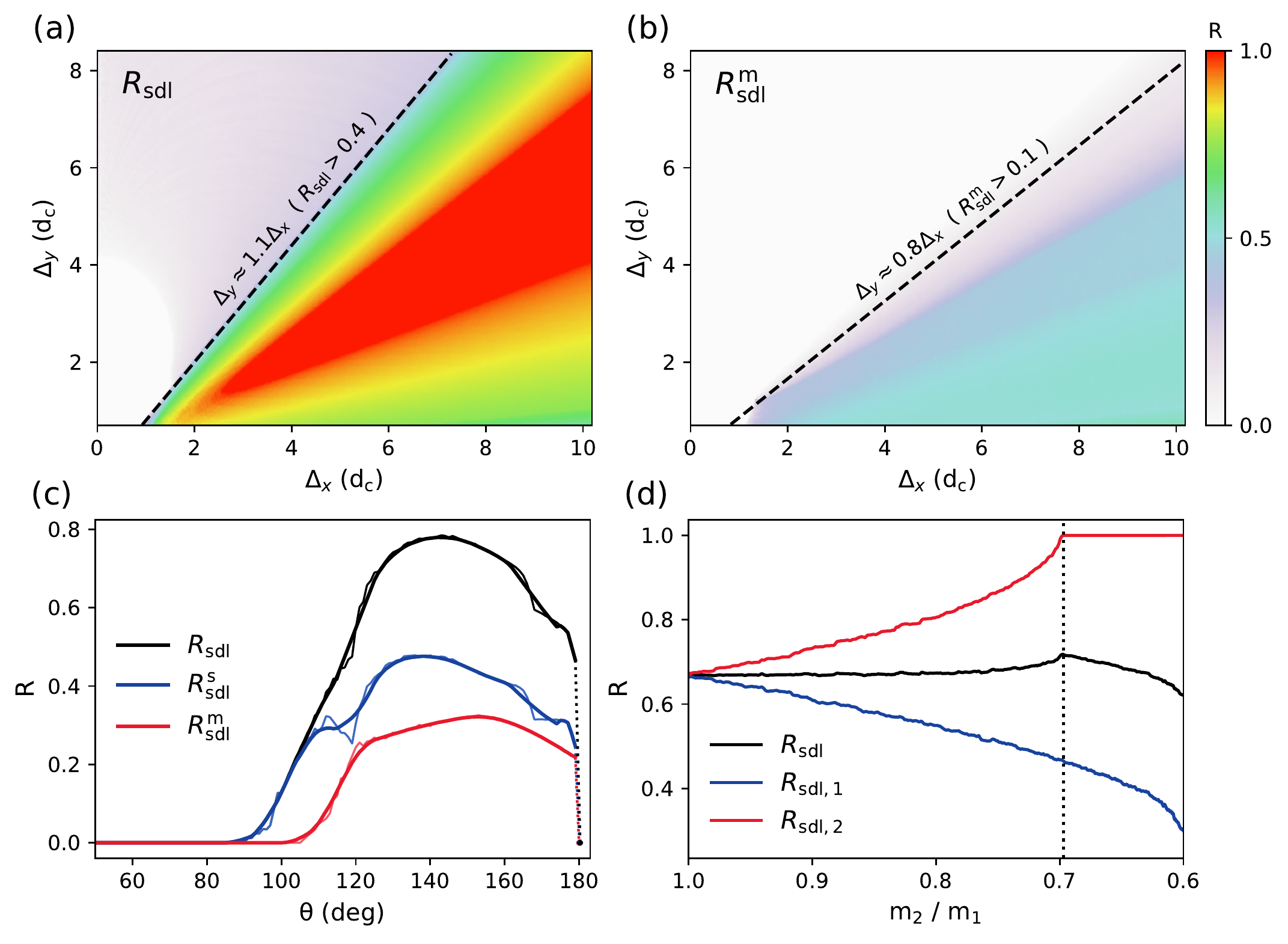}
\caption{Occurrence rate of saddle-like decay-index profiles in quadrupolar configurations. (a)-(b) The distribution of $R_\mathrm{sdl}$ and $R_\mathrm{sdl}^\mathrm{m}$ in the $\Delta_x-\Delta_y$ space. $\Delta_x$ and $\Delta_y$ denote the displacement of BP2 with respect to BP1 in the $X$ and $Y$ direction (cf. Figure~\ref{fig:sim_quadrupole}), respectively. (c) $R_\mathrm{sdl}$ as the function of the angle $\theta$ between two dipoles. Blue (red) curve denotes $R_\mathrm{sdl}^\mathrm{s}$ ($R_\mathrm{sdl}^\mathrm{m}$) after applying a 10-point smoothing procedure. Dashed curves show corresponding raw values. (d) $R_\mathrm{sdl}$ as function of the ratio between dipole magnitude $m_2/m_1$. Blue (red) curves denotes $R_\mathrm{sdl,1}$ ($R_\mathrm{sdl,2}$) above the PIL segment that is closer to BP1 (BP2). \label{fig:sim_diff}}
\end{figure*}

\begin{figure*}
\centering
\includegraphics[width=0.95\linewidth]{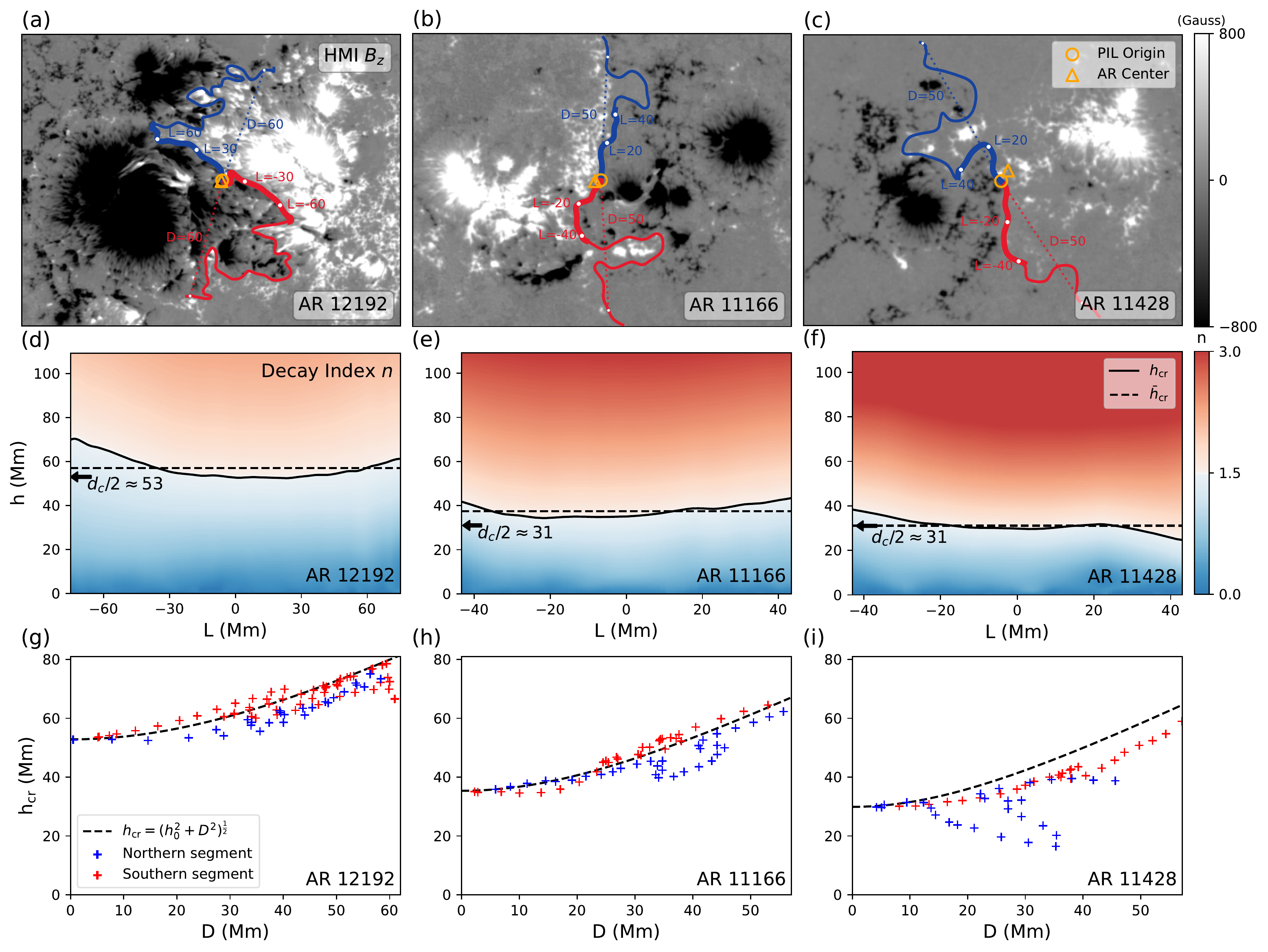}
\caption{Characteristics of critical height $h_\mathrm{cr}$ in three representative bipolar ARs, 12192, 11166, and 11428. (a)-(c) HMI $B_z$ maps saturated at $\pm800$ G. The half centroid distance $d_c/2$ in three ARs is around 53, 31, and 31~Mm, respectively. The yellow triangles denote the AR centers, the midpoint of the centroids of opposite polarities. The yellow circles mark the PIL origin, the point closest to the AR center  (see \S\ref{subsubsec:dipoleobs}). The PIL segment to the north (south) of the origin is shown in blue (red). (d)-(f) 2D map of decay index $n$ as a function of height $h$ and the arc length $L$ from the PIL origin. The range of $L$ is denoted by thick PILs in panels (a)-(c). Solid curves denote $h_\mathrm{cr}$. Horizontal dashed lines mark $\bar{h}_\mathrm{cr}$ (Eq.~\ref{eq:dip_av_hcrit}), which is close to $d_c/2$. (g)-(i) $h_\mathrm{cr}$ as a function of the linear distance $D$ from the AR center above the entire PIL. Dashed curves show $h_\mathrm{cr}$ for a dipole field Eq.~\ref{eq:dip_hcrit_original}, with $h_0$ given by the critical height at the PIL origin obtained in the potential field. \label{fig:obs_dipole}}
\end{figure*}

\begin{figure*}
\centering
\includegraphics[width=0.95\linewidth]{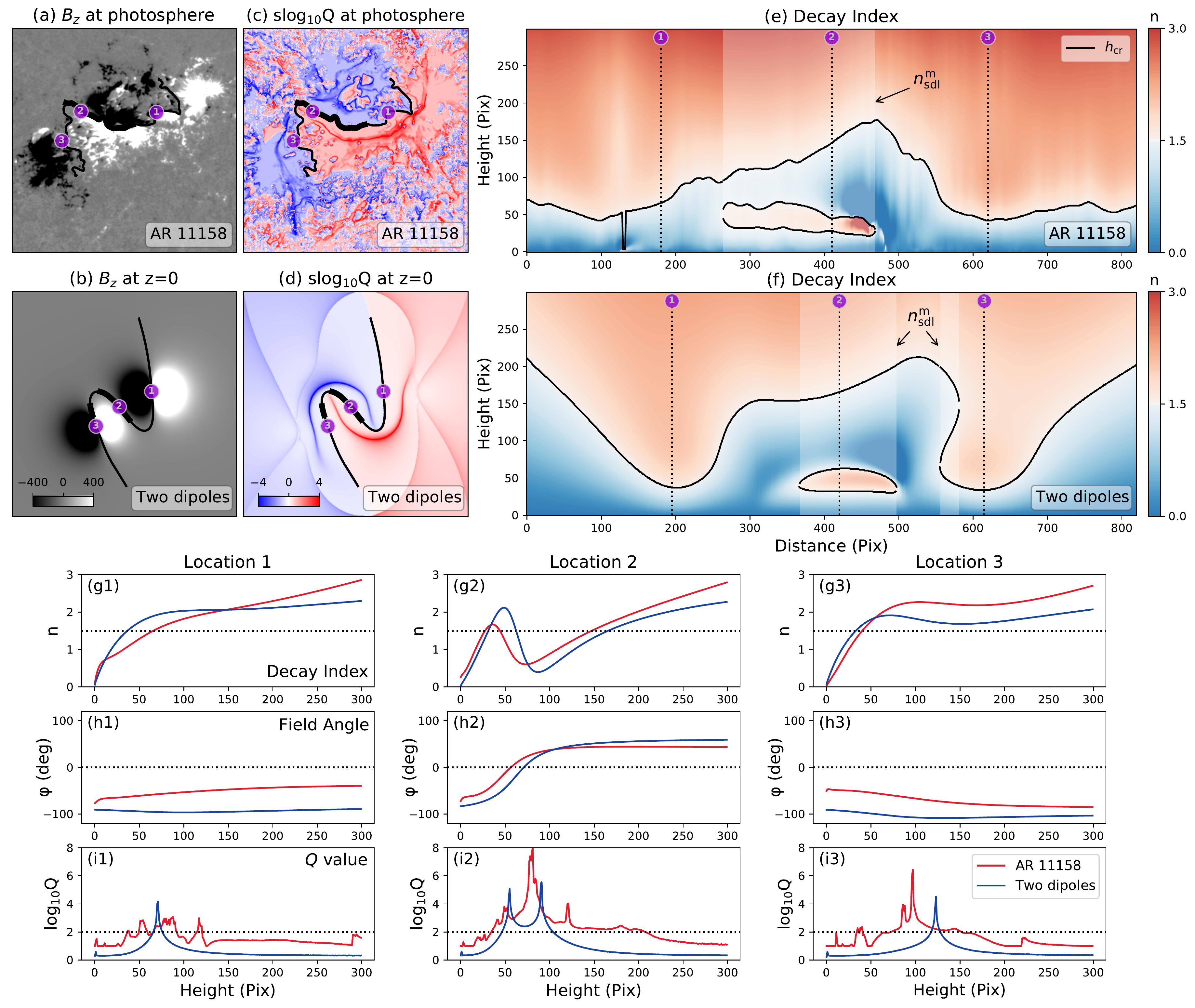}
\caption{Background field of NOAA AR 11158 approximated by potential extrapolation vs by two dipoles. (a)-(b) Photospheric flux distributions of AR 11158 and corresponding quadrupole field. Black curves denote the PIL. The $n_\mathrm{sdl}^\mathrm{m}(h)$ profile appears above the PIL segments as denoted by thicker curves. (c)-(d) Corresponding map of signed $\log_{10}Q$, i.e., $\mathrm{slog}~Q \equiv \mathrm{sgn}(B_z)\times\log_{10}Q$. (e)-(f) 2D maps of decay index $n$ as a function of height and arc length along the PIL, in units of pixels.  $h_\mathrm{cr}$ is marked by black solid curves. Lightly shaded regions indicate the presence of $n_\mathrm{sdl}^\mathrm{m}(h)$. At three selected locations, we show the height variation of $n$ (g1--g3), the angle $\varphi$ between the transverse field component and the PIL (h1--h3), and $\log_{10}Q$ (i1--i3), respectively, in the potential (red) and quadrupole field (blue). Horizontal dotted lines mark the torus instability threshold ($n=1.5$), field reversal location ($\varphi=0$), and quasi-separatrix layers ($\log_{10}Q\ge2$), respectively. \label{fig:obs_11158}}
\end{figure*}

\begin{figure*}
\centering
\includegraphics[width=0.85\linewidth]{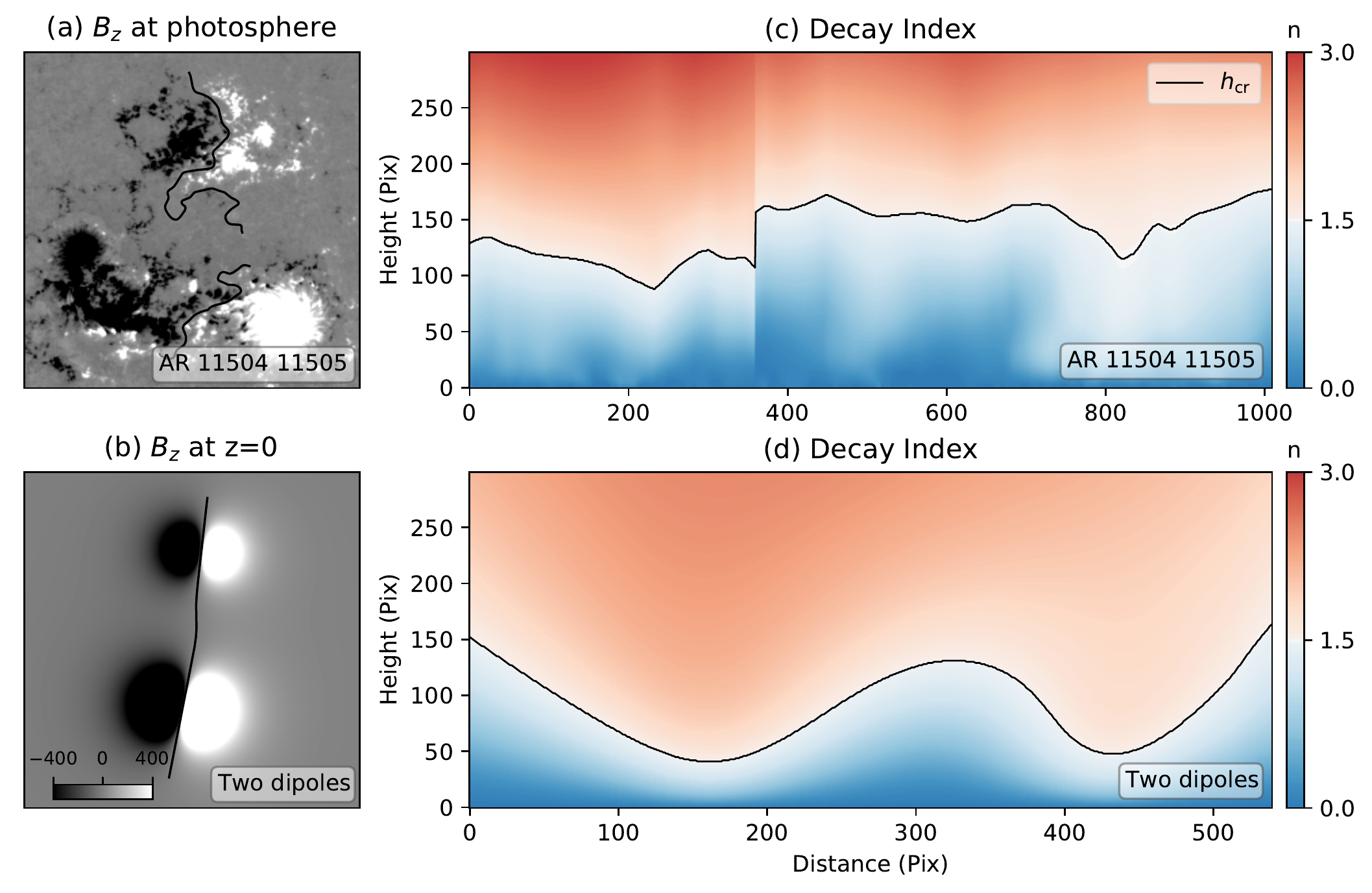}
\caption{Background field of NOAA ARs 11504/11505 approximated by potential extrapolation vs by two dipoles. The format is similar to Figure~\ref{fig:obs_11158}(a, b, e, and f).\label{fig:obs_11504_11505}}
\end{figure*}

\begin{figure*}
\centering
\includegraphics[width=0.95\linewidth]{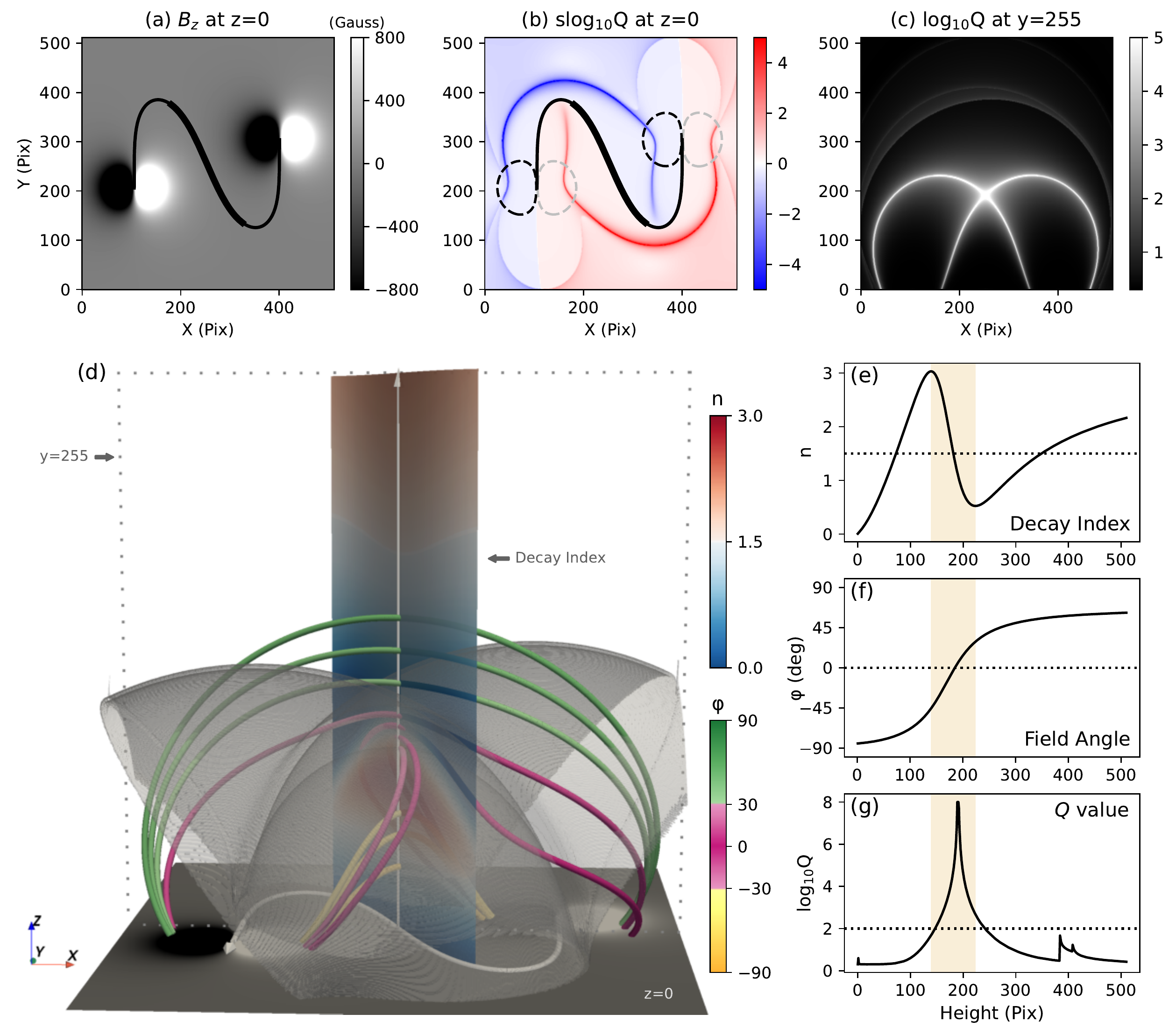}
\caption{Association of saddle-like decay index profiles with magnetic skeletons. (a) The pseudo photospheric flux distribution of the quadrupolar field. (b) Map of signed $\mathrm{log}~Q$ in the pseudo photosphere. Dashed contours indicate two sunspot pairs. Solid line denotes the PIL. $n_\mathrm{sdl}^\mathrm{m}(h)$ profiles appear above the PIL segment indicated by thickened curves. (c) Logarithmic $Q$ in the cutting plane $y=255$, which is denoted by the vertical dashed rectangle in (d). (d) 3-D isosurfaces of $\mathrm{log_{10}}~Q=3.5$ with representative field lines. The color of each individual field lines corresponds to the angle between $\mathbf{B_t}$ and the PIL direction in the center. The decay index in the cross section is color coded. (e)-(g) Height variation of $n$, $\varphi$, and $\mathrm{log_{10}}~Q$ along the vertical direction in the center of the cross section in (d). Horizontal dotted lines mark the torus instability threshold ($n=1.5$), field reversal location ($\varphi=0$), and quasi-separatrix layers ($\log_{10}Q\ge2$), respectively. Yellow shades denote the region between the saddle top and bottom. An animation of panel (d) in different perspectives is available online. The video duration is 9 s. \label{fig:Sdl&HFT}}
\end{figure*}

To summarize, our study corroborates that the critical height is roughly half of the centroid distance between opposite polarities in a dipole field and clarifies where a saddle-like decay-index profile can be found in a quadrupole field. With some caution, these results can be applied to bipolar and quadupolar ARs, respectively. Our study also reveals that saddle-like profiles may be associated with magnetic skeletons, such as a null point (cf. Fig.~\ref{fig:sim_quadrupole}(b \& g)) or an HFT (Fig.~\ref{fig:Sdl&HFT}), which is anticipated to serve as a ``double-edged sword'' in eruptive processes in case magnetic reconnection with the eruptive structure taking place at such skeletons: reconnection may open up the overlying field, facilitating the eruption, but at the same time shed the flux of the eruptive structure, causing its disintegration or even failed eruption. Thus, the zone of stability at the saddle may indeed provide some constraint for the eruptive structure. This may explain the mixing results obtained in observational studies, as mentioned in the beginning of \S\ref{subsec:role_of_sdl}.

\acknowledgements This work was supported by NSFC grants 41774150, 11925302 and 42188101 and the Strategic Priority Program of the Chinese Academy of Sciences (XDB41030100). The authors thank the HMI consortia for excellent data.

\bibliography{sample63}{}
\bibliographystyle{aasjournal}

\newpage

\listofchanges
\end{document}